\documentclass[a4paper,
twocolumn,preprintnumbers,citeautoscript,newabstract,aps,superscriptaddress]{revtex4-2} 
\usepackage{amsmath}
\usepackage{amsfonts}
\usepackage{dsfont}
\usepackage{amssymb}
\usepackage{slashed}
\usepackage[hidelinks=true]{hyperref}
\usepackage{graphicx}
\usepackage{xcolor}
\usepackage{bbm}

\renewcommand{\bar}[1]{\overline{#1}}

\newcommand{\mat}[1]{\left(\begin{matrix}#1\end{matrix}\right)}

\newcommand{\I}{\ensuremath{\mathrm{i}}}
\newcommand{\e}{\ensuremath{\mathrm{e}}}

\definecolor{darkgreen}{HTML}{109930}

\begin{document}
\title{Goofy transformations and the hierarchy problem}
\author{Andreas Trautner}\email[]{trautner@cftp.ist.utl.pt}
\affiliation{CFTP, Departamento de F\'isica, Instituto Superior T\'ecnico, Universidade de Lisboa, Avenida Rovisco Pais 1, 1049 Lisboa, Portugal}

\begin{abstract}\noindent
Goofy transformations of the Standard Model (SM) Higgs field generally prohibit its bare mass term. This opens up an entirely new class of solutions to the electroweak (EW) hierarchy problem. We argue that these can be intrinsically linked to the flavor structure and origin of CP violation.
\end{abstract}
\maketitle

\widowpenalty1000000\clubpenalty1000000
\textit{Introduction.--}
Starting with Ferreira, Grzadkowski, Ogreid, and Osland~(FGOO)~\cite{Ferreira:2023dke}, there recently has been some activity to explore new, so far unknown possible global symmetry transformations with unusual properties~\cite{Haber:2025cbb,Trautner:2025yxz,Ferreira:2025ate}. As contrasted to \textit{regular} symmetry transformations, the new, so-called \textit{goofy} transformations can be defined by their unusual feature of non-trivially transforming (gauge-)kinetic terms of a quantum field theory~\cite{Trautner:2025yxz}. 

Since (gauge-)kinetic terms and mass terms are typically constructed from the same bi-linear invariant contractions over global field space, it is a direct consequence of the goofy nature of a transformation that it will also act non-trivially on possible mass terms. Specifically, this applies to bare scalar mass terms, for what reason it was proposed in~\cite{Trautner:2025yxz} that goofy transformations can give rise to a new class of solutions to the electroweak hierarchy problem. 

Controlling the bare scalar mass term by a symmetry is an essential feature of solutions to the hierarchy problem such as supersymmetry (which links the bare scalar mass terms to fermion masses that can be controlled by chiral symmetry), composite Higgs models (where the Higgs is a pseudo-Nambu-Goldstone boson with bare mass prohibited by shift symmetry), and also conformal symmetry (which, however, is believed to be broken homogeneously by all dimensionful parameters). Goofy transformations take their own spot in this space of possible solutions.

All of these solutions require a breaking of the protecting symmetry in order to be phenomenologically viable in the sense that they need to allow for electroweak symmetry breaking and the observed Higgs mass. The mechanism of breaking typically renders the Higgs mass computable and introduces deviations in Higgs phenomenology as compared to the plain SM. 
Also in goofy situations phenomenological consequences are computable in specific models and will lead to testable predictions at future colliders and Higgs factories.

\enlargethispage{1cm}
Exploration of phenomenological consequences, however, is beyond the scope of this work. We will merely outline the most basic idea by stating explicit classes of goofy transformations which can prohibit the scalar quadratic term, while allowing fermion Yukawa couplings. We will illustrate how the Higgs mass can be re-introduced by a spurion argument. 

Irrespective of whether or not one approves of the concept of goofy transformations, they factually lead the way to a previously unknown class of all-order fixed points of the renormalization group~(RG) flow~\cite{Ferreira:2023dke,Trautner:2025yxz}. Such an all-order stable behavior is well-known usually only for \textit{regular} symmetry transformations, as commonly accepted based on 't~Hooft's technical naturalness argument~\cite{tHooft:1979rat}. It turns out that also the all-order stability of goofy transformations can be understood by an extension of this argument, and the insight that the symmetry of the beta functions can be larger than the symmetry of the action~\cite{Trautner:2025yxz,deBoer:2025xx}.

As a model building choice, one possibility to 
compensate the unusual transformation behavior of gauge-kinetic terms under goofy transformations is by introducing an equally goofy transformation of space-time and gauge fields~\cite{Ferreira:2023dke},
\begin{align}\label{eq:goofy_st}
&\partial_\mu\,\mapsto\,-\I\partial_\mu\;,&
&\text{and}&
&A_\mu\,\mapsto\,-\I A_\mu\;.&
\end{align}
The combined transformation then acts as a regular symmetry, but only leaves invariant the full effective potential if one simultaneously performs a sign flip of the squared UV cutoff, or squared $\bar{\mathrm{MS}}$ renormalization scale~\cite{Grzadkowski:2024, Ferreira:2025ate} (see also \cite{Cao:2023kgq,Pilaftsis:2024uub}). This might be important to take into account for the ``intrinsic'' hierarchy problem of the SM~\cite{Wells:2025hur}, as a spurious transformation behavior of the cutoff under goofy transformations can determine whether or not, and at what order, it contributes to the otherwise prohibited Higgs quadratic term.

An alternative route to understand the all-order RG stability of the goofy parameter relations -- without imposing the exotic space-time transformations in~\eqref{eq:goofy_st} --  stems from the observation that the breaking of goofy symmetry by the gauge-kinetic terms \textit{can} be \textit{soft}, in the sense that it does not enter quantum corrections of other goofy non-invariant operators, which also explains their RG stability to all orders~\cite{Trautner:2025yxz}. Whether or not the explicit breaking by the gauge-kinetic terms actually \textit{is} soft for a specific goofy transformation, so far has been determined on a case by case basis using explicit higher-loop computations or the all-order arguments of~\cite{Trautner:2025yxz,deBoer:2025xx}. Some generally applicable arguments are given in footnote~\footnote{\label{foot:softornot}
The gauge-kinetic terms are dimension-four, naively marginal operators, hence, might be expected to lead to a hard symmetry breaking. However, it has been shown in~\cite{Trautner:2025yxz} that this is not necessarily the case. Specifically, it was empirically found in the 2HDM that cases where \textit{all} gauge-kinetic terms break the goofy transformation (``global-sign-flipping goofy transformations'') lead to radiatively stable goofy parameter conditions (i.e.~the global sing-flipping gauge-kinetic terms are only softly breaking the goofy symmetry), while ``relatively-sign-flipping'' goofy transformations of the gauge-kinetic terms lead to a hard breaking of the goofy transformations (i.e.~the goofy parameter conditions are then not radiatively stable). On the one hand, these statements straightforwardly generalize to other models also including fermion gauge-kinetic terms, at least if the new particles share a gauge interactions with the scalars. On the other hand, the statements about radiative stability for global-sign-flipping transformations break down if there appear  goofy-invariant trilinear couplings (such as, for example, a trilinear scalar coupling). A generally applicable theorem for arbitrary models seems possible but has not been formulated.}.
Unlike in the case of additional space-time symmetries, in this picture, nothing has to be imposed about the transformation behavior of the RG scale or cutoff, and physical scales can exist which do not have the correct spurion transformation behavior to enter the Higgs quadratic term, hence, giving way for a stabilization of a hierarchy of scales.

\textit{Goofy transformations.--}
We denote the SM Higgs doublet $H$ and its conjugate doublet $\widetilde H:=\varepsilon H^*$. The discussion is completely analogous to the one given for the two-Higgs-doublet model in~\cite{Trautner:2025yxz}, however, without the additional complication of ``Higgs flavor'' space.
Starting from a canonical kinetic term and keeping it hermitean, the most general possible goofy transformations of the Higgs field are given by~\footnote{%
Skeptics of such a transformation may recall that a field and its hermitean conjugate are necessarily \textit{independent} degrees of freedom of a QFT. Alternatively, one may also spell out everything in real fields which is to various level of detail discussed in~\cite{Ferreira:2023dke,Haber:2025cbb,Trautner:2025yxz} but does not give any more insights than keeping the notation compact.}
\begin{align}\label{eq:RegularTrafos}
    &\mat{H \\ \widetilde H}\mapsto\mat{  \e^{\I\alpha}&  0\\   0&  -\e^{-\I\alpha}}
    \mat{H \\ \widetilde H}\,,\quad\text{or}& \\ \label{eq:CPTrafos}  
    &\mat{H \\ \widetilde H}\mapsto\mat{  0&  \e^{\I\beta}\\   -\e^{-\I\beta}&  0}
    \mat{H \\ \widetilde H}\,,&
\end{align}
for flavor- or CP-type transformations, respectively~\footnote{%
\label{fot:CP}%
It is left implicit throughout this work, but for CP-type transformation we always additionally perform the transformation $(t,x)\mapsto(t,-x)$ as well as the usual action of the complex conjugation outer automorphism on gauge fields and fermions~\cite{Grimus:1995zi,Trautner:2016ezn}. For fermions the CP transformation on flavor and spinor indices reads, for example, $Q_\mathrm{L}\mapsto C_{Q_\mathrm{L}}\,\mathcal{C}\,Q_\mathrm{L}^*$, where $\mathcal{C}$ is the charge conjugation matrix and $C_{Q_\mathrm{L}}$ a possible flavor-space transformation.}.  The phases $\alpha$ and $\beta$ here can always be modified by performing an additional global hypercharge transformation and we often adopt a convention such as to set $\alpha=\beta=0$ for the Higgs transformation~\footnote{%
We remark that an independent passive field redefinition of $H$ and $\protect\widetilde{H}$ cannot be used to absorb the phases in Eq.~\eqref{eq:RegularTrafos} since such an operation commutes with this transformation. Passive field redefinitions, however, can be used to 
absorb the phases of the CP-type transformation in Eq.~\eqref{eq:CPTrafos}, which can be used to show that it is basis-change equivalent to an order~$2$ transformation. Moreover, passive ``goofy'' field redefinitions can be used to change to a basis with different sign for the kinetic term.}.

The Higgs bilinear and \mbox{gauge-kinetic} terms transform with a sign flip~\footnote{%
We should remark that a wrong signed kinetic term here is related by an unphysical field redefinition to a theory with correct-signed kinetic term, illustrating that the sign of the kinetic term alone does not determine physicality of a theory.}
\begin{align}
\left(D^\mu H\right)^\dagger\left(D_\mu H\right)~&\mapsto~
-\left(D^\mu H\right)^\dagger\left(D_\mu H\right)\;,\\
\mu_H^2\,H^\dagger H~&\mapsto~-\mu_H^2\,H^\dagger H\;.
\end{align}
This shows that both terms explicitly break the goofy transformations.

The Higgs also has Yukawa couplings to fermions. For the quark sector those read
\begin{align}\label{eq:SMYukawas}\notag
 -\mathcal{L}_\mathrm{Yuk.}~=&~\bar{Q}_\mathrm{L}\widetilde{H}\,Y_u\,u_\mathrm{R}+\bar{Q}_\mathrm{L}H\,Y_d\,d_\mathrm{R}+ & \\
 &~\bar{u}_\mathrm{R}\widetilde{H}^\dagger\,Y_u^\dagger\,Q_\mathrm{L}+\bar{d}_\mathrm{R}H^\dagger\,Y_d^\dagger\,Q_\mathrm{L}.& 
\end{align}
Without imposing additional requirements, the flavor space coupling matrices $Y_u$ and $Y_d$ explicitly break the goofy transformations. This can be prevented by extending the goofy Higgs transformations to fermions.   Requesting invariant Yukawa couplings unavoidably requires some form of goofy transformation also for the fermions~\footnote{%
This is easy to see by the fact that if a coupling $H\mathcal{O}$ is invariant, then also $\mathcal{O}^\dagger H^\dagger$ is present. If the Higgs transforms goofy, also $\mathcal{O}$ has to transform goofy.  
}. Extending the goofy transformations to flavor is an interesting avenue for model building. While the full exploration of all possibilities is important, here we will focus on one thing only: showing that a goofy transformation exists that prohibits the Higgs quadratic term but allows $Y_u$ and $Y_d$, ensuring that we can reproduce the correct SM flavor phenomenology. 

Extending goofy transformations to fermions, there is a qualification necessary, depending on if one requests the  additional space-time symmetries \eqref{eq:goofy_st} or not. For the scalar kinetic term, the requirement of hermiticity in the case of no additional space-time transformation imposes the very same requirement as demanding invariance in the presence of the additional space-time transformations, Eq.~\eqref{eq:goofy_st}.
By contrast, for the fermion kinetic term, these two requirements do not align. For example, consider a generic fermion kinetic term (eventually we will take $\Psi=Q_\mathrm{L},\,u_\mathrm{R},\,d_\mathrm{R}$)
\begin{equation}
\mathcal{L}_\mathrm{\Psi}~=~
\I\,\bar{\Psi}\,\slashed{D}\,\Psi~=~
\I\,\bar{\Psi}\,\gamma^\mu\left(\partial_\mu-\I\,g\,A_\mu \right) \Psi\;.
\end{equation}
Under the most general possible unitary flavor- or CP-type transformation (commuting with gauge symmetries) the fermions transform as
\begin{align}\label{eq:RegularTrafosFermions}
    &\mat{\Psi^{\phantom{*}}\! \\ \Psi^*\!}\mapsto\mat{  A_{\Psi}&  0\\   0& B_{\Psi}^*}
    \mat{\Psi^{\phantom{*}}\! \\ \Psi^*\!}\,,\quad\text{or}& \\   
    &\mat{\Psi^{\phantom{*}}\! \\ \Psi^*\!}\mapsto\mat{  0&  C_{\Psi}\\   D^*_{\Psi}&  0}
    \mat{\Psi^{\phantom{*}}\! \\ \Psi^*\!}\,,&
\end{align}
where we have included $3\times3$ unitary transformation matrices in flavor space $A_\Psi$, $B_\Psi$, $C_\Psi$, and~$D_\Psi$~\footnote{%
See~\cite{Trautner:2025yxz} for details in the analogous case of scalars and keep in mind footnote~\cite{Note3}.}.

Starting from the canonical basis, and without additional exotic space-time transformations, hermiticity of the kinetic term requires $B_\Psi^\dagger A_\Psi=\pm\mathbbm{1}$ or $D_\Psi^\dagger C_\Psi=\pm\mathbbm{1}$ for flavor universal regular $(+)$ or goofy $(-)$ transformations, respectively. Different transformations for different $\Psi$'s, and also flavor non-universal transformations are possible. By contrast, superimposing the additional exotic space-time symmetry \eqref{eq:goofy_st}, 
the overall invariance of the kinetic term requires $B_\Psi^\dagger A_\Psi=\I\mathbbm{1}$ or $D_\Psi^\dagger C_\Psi=\I\mathbbm{1}$, which in this case has to hold \textit{universally} for all $\Psi=Q_\mathrm{L},\,u_\mathrm{R},\,d_\mathrm{R}$. Obviously, these are very different constraints that can lead to different possibilities regarding the possibly allowed transformations and the resulting phenomenology. 
 
For the case of no extra space-time symmetry there is a lot of model building freedom. In particular, which of the fermion kinetic terms transforms non-trivially and, therefore, explicitly breaks a given goofy transformation, depends on the precise charge assignments under the transformation.

For example, choosing the convention $\alpha=0$ in Eq.~\eqref{eq:RegularTrafos} and keeping $Q_\mathrm{L}$ inert, the simple fermion transformation 
\begin{equation}
 \mat{u_\mathrm{R}, u^*_\mathrm{R}}~\mapsto~\mat{-u_\mathrm{R}, u^*_\mathrm{R}}\,,\quad \mat{d_\mathrm{R}, d^*_\mathrm{R}}~\mapsto~\mat{d_\mathrm{R}, -d^*_\mathrm{R}}
\end{equation}
can ensure invariant Yukawa couplings, while the Higgs mass term is prohibited. The transformation
is explicitly broken by the $H$, $u_\mathrm{R}$ and $d_\mathrm{R}$ kinetic terms. The explicit breaking by kinetic terms is not automatically bad, since such a breaking \textit{can} be soft, see~\cite{Trautner:2025yxz} and footnote~\cite{Note1}.
Irrespective of whether the breaking is hard or soft,  the goofy transformation and its breaking control radiative corrections in potentially phenomenologically interesting ways~\cite{Trautner:2025yxz}.

For the CP-type transformations, of course, all fermions have to transform. This does not, however, mean that they all should transform goofy. Adopting the phase convention $\beta=0$ in Eq.~\eqref{eq:CPTrafos} and
letting $Q_\mathrm{L}$ transform in the canonical \textit{regular} way 
$(Q_\mathrm{L},Q_\mathrm{L}^*)\mapsto(Q_\mathrm{L}^*,Q_\mathrm{L})$, c.f.~\cite{Note2}, 
the simple CP-type goofy transformation 
\begin{equation}
 \mat{u_\mathrm{R}, u^*_\mathrm{R}}~\mapsto~\mat{-u^*_\mathrm{R}, u_\mathrm{R}}\,,\; \mat{d_\mathrm{R}, d^*_\mathrm{R}}~\mapsto~\mat{d^*_\mathrm{R}, -d_\mathrm{R}}
\end{equation}
can ensure prohibition of the Higgs mass term while
transforming the Yukawa couplings in the canonical way. This transformation is explicitly broken  by the  gauge-kinetic terms of $H$, $u_\mathrm{R}$, $d_\mathrm{R}$ and the CP non-invariant part of the Yukawa couplings $Y_{u,d}-Y^*_{u,d}$. This shows a connection between the Higgs mass term and potential sources of CP violation.

Another choice of flavor universal goofy transformation is the simultaneous sign flip of all fermion gauge-kinetic terms. In this case the fermion transformations for all $\Psi=Q_\mathrm{L},u_\mathrm{R}, d_\mathrm{R}$ reads
\begin{align}
 &(\Psi,\Psi^*)\mapsto\left(\e^{\I\alpha_\Psi}\Psi,-\e^{-\I\alpha_\Psi}\Psi^*\right)\;,\quad\mathrm{or}& \\
 &(\Psi,\Psi^*)\mapsto\left(\e^{\I\beta_\Psi}\Psi^*,-\e^{-\I\beta_\Psi}\Psi\right)\;,&
\end{align}
for flavor or CP-type transformations respectively. It turns out that for these transformations there is no consistent possible choice of phases $\alpha$ and $\alpha_{Q_\mathrm{L},u_\mathrm{R},d_\mathrm{R}}$, 
or $\beta$ and $\beta_{Q_\mathrm{L},u_\mathrm{R},d_\mathrm{R}}$ such as to allow the Yukawa couplings. In this sense, goofy transformations are required to be chiral in order to \textit{allow} the Yukawa couplings.

One more interesting possibility for a flavor-type goofy transformation could be one where in addition to the Higgs field (conventions $\alpha=0$) only $Q_\mathrm{L}$ transforms. For the flavor universal case this means 
\begin{equation}
 \mat{Q_\mathrm{L}, Q^*_\mathrm{L}}~\mapsto~\mat{\e^{\I\alpha_{Q_\mathrm{L}}}Q_\mathrm{L}, -\e^{\I\alpha_{Q_\mathrm{L}}} Q^*_\mathrm{L}}\,.
\end{equation}
There are two possible choices of phase ($\alpha_{Q_\mathrm{L}}=0,\pi$) to \textit{either} allow $Y_u$ \textit{or} $Y_d$, but not both. This gives an excellent starting point to explain the suppression of the bottom-type Yukawas. Even though the goofy  transformation is explicitly broken by the $H$ and $Q_\mathrm{L}$ gauge-kinetic terms, we do not expect radiative corrections to re-introduce $Y_d$ here, simply because the vanishing Yukawa couplings are additionally protected by chiral symmetry (this does not preclude finite contributions to $Y_d$ in SM extensions). 

We also consider the analogous CP-type transformation ($\beta=0$), 
\begin{equation}
 \mat{Q_\mathrm{L}, Q^*_\mathrm{L}}~\mapsto~\pm\mat{Q_\mathrm{L}^*, -Q_\mathrm{L}}\,,
\end{equation}
with $u_\mathrm{R}$ and $d_\mathrm{R}$ inert. This transformation can be preserved if $Y_u=\pm Y_u^*$ and $Y_d=\mp Y_d^*$ (all signs correlated). This would demand purely imaginary Yukawa couplings for one type of quarks, and is consistent with CP conservation. 

All these examples show that goofy transformations which preserve the hermiticity of the gauge-kinetic terms without invoking additional space-time symmetries,
are interesting, not only because they can prohibit the Higgs quadratic, but also because they necessarily have to have some level of flavor dependence. 

Let us also consider the case of imposing the additional space-time transformation, Eq.~\eqref{eq:goofy_st}. If this is combined with the transformations \eqref{eq:RegularTrafos} or \eqref{eq:CPTrafos} and demanding invariance of the fermion kinetic terms, the flavor universal fermion goofy transformations are limited to 
\begin{align}
 &(\Psi,\Psi^*)\mapsto\left(\e^{\I\alpha_\Psi}\Psi,\I\e^{-\I\alpha_\Psi}\Psi^*\right)\;,\quad\mathrm{or}& \\
 &(\Psi,\Psi^*)\mapsto\left(\e^{\I\beta_\Psi}\Psi^*,\I\e^{-\I\beta_\Psi}\Psi\right)\;,&
\end{align}
for flavor or CP-type transformations, respectively. This always necessarily acts simultaneous on all $\Psi=Q_\mathrm{L},u_\mathrm{R}, d_\mathrm{R}$.
The additional requirement of invariant Yukawas fixes the relative phases to 
\begin{align}
 \alpha_{Q_\mathrm{L}}-\alpha_{u_\mathrm{R}}~&=~-\frac{\pi}{2}-\alpha\;,\\
 \alpha_{Q_\mathrm{L}}-\alpha_{d_\mathrm{R}} ~&=~\frac{\pi}{2}+\alpha\;,
\end{align}
or 
\begin{align}
 \beta_{Q_\mathrm{L}}-\beta_{u_\mathrm{R}}~&=~-\frac{\pi}{2}-\beta\;, \\
 \beta_{Q_\mathrm{L}}-\beta_{d_\mathrm{R}} ~&=~\frac{\pi}{2}+\beta\;, 
\end{align}
and $Y_{u,d}=Y_{u,d}^*$ in the case of the CP transformation. 
Phase conventions can be chosen, including adding in an overall global hypercharge and/or baryon number transformation~\footnote{%
Taking $\alpha=\pi/2$ and making the choice $\alpha_{Q_\mathrm{L}}=\pi/4$, our conditions require  $\alpha_{u_\mathrm{R}}=\alpha_{d_\mathrm{R}}=5\pi/4$ which is consistent with the special case considered in~\cite{deBoer:2025jhc}.}.

\textit{Discussion.--}
All of our examples show how goofy transformations can be invoked to prohibit the Higgs quadratic term while permitting SM Yukawa couplings. Even if a goofy transformation is explicitly broken by kinetic terms, it \textit{can} give rise to all-order RG stable fixed points, including mass terms (see footnote~\cite{Note1}). This has been explicitly shown for some cases of the 2HDM in~\cite{Ferreira:2023dke, Trautner:2025yxz}, where RG stable fixed points exist even if the canonical gauge-kinetic terms explicitly (softly) break the corresponding goofy transformation~\footnote{%
We add to~\cite{Trautner:2025yxz} that for all cases with exact protection of mass terms, one can always find a full rank flavor non-universal (hence non-canonical) kinetic term (this specifies a different theory!) which is actually conserved under a (in that case Higgs flavor dependent) goofy transformation.
This implies that the theory with invariant but non-canonical kinetic term, is actually a non-goofy theory - as the kinetic term is invariant - just written in a non-canonical basis. Clearly a (non-)invariant transformation behavior of any term in an irrep is a basis invariant feature.}. Irrespective of whether or not a goofy transformation is actually imposed, the transformation acts as an outer automorphism and, therefore, provides a sufficient condition for a symmetry of the system of beta functions that can be larger than the symmetries of the effective action~\cite{Trautner:2025yxz, deBoer:2025xx}.  

This implies that a vanishing beta function of $\mu_H$ can be stabilized to all order by imposing a goofy transformation of the SM Higgs field. Of course, in the pure SM this conclusion seems somewhat trivial, as $\mu_H$ is the only dimensionful parameter.
The true power of goofy transformations is that they 
will also dictate the possible (potentially finite) corrections to $\mu_H$ in case other mass scales are present in extensions of the SM or taking into account gravity, i.e.\ help in addressing the ``extrinsic'' hierarchy problem of the SM~\cite{Wells:2025hur}.

While goofy transformations are instrumental to prohibit the Higgs quadratic term to begin with, we stress that they do not seem to dictate a unique route along which the Higgs mass and EW scale is re-introduced. In this sense, the possible model building landscape may be as vast as, for example, in supersymmetric solutions to the hierarchy problem. Hence, it is not possible to explore the full landscape of possibilities in the course of just this letter. Rather, we would like to close by giving a few universal arguments that will hold true irrespective of the concrete scenario.

From a simple diagrammatic argument, we expect that finite contributions to $\mu_H$ can arise only if the corresponding scale can, in principle (i.e.\ before imposing goofy parameter relations), enter the beta function of~$\mu_H$. Once a goofy transformation is imposed to prohibit $\mu_H$ and its associated counter-term, the RG corrections must vanish, which does not automatically preclude finite contributions to vanish (finite contributions have been explicitly computed in~\cite{Pilaftsis:2024uub}). Finite contributions to the Higgs mass can be computed from the full effective potential. The Coleman-Weinberg dynamical generation of scale here is also controlled by goofy transformations, which can be especially interesting taking into account their flavor non-universal variants.

We will not discuss UV completions here, but use a spurion argument about new scales to show how goofy constraints play a role. Regarding the Higgs mass term, naive dimensional power counting suggests that any scale should enter the beta functions. However, whether or not it actually does, depends on whether its contribution can be supported by corresponding covariant transforming combination of other couplings (see~\cite{Trautner:2025yxz}). For example, a new high scale that transforms invariant under goofy transformations can affect the Higgs mass term only if an appropriate combination of other couplings -- necessarily goofy-odd in combination -- allows to support the contribution to the beta function. This implies that once the goofy symmetry is imposed, such a combination of couplings is absent and the Higgs mass insensitive to the new high scale. 
However, things are more involved, as we would like to keep the kinetic terms intact, which are necessarily explicitly goofy-breaking at least for the Higgs field.
As we stressed already above, this does not automatically mean the sensitivity to the high scale is restored, as soft symmetry breaking by kinetic terms is a possibility that has been proven to exist. 

Another possibility is that we have a high scale spurion that transforms odd under the goofy transformation.
This can be the case, for example, if a new scalar field 
that itself transforms goofy-odd condenses to spontaneously break the goofy transformation. Such a spontaneous breaking can be triggered by a goofy-even mass scale, or via dimensional transmutation \'a la Coleman-Weinberg. Such a spurion could couple to the Higgs mass term potentially already at the tree level, as the portal term of the new scalar to the Higgs would be allowed by the goofy transformation. This would render the Higgs mass computable (i.e.\ still prohibit the Higgs mass counter-term), but suggests that the scale of spontaneous goofy breaking should not be too far away from the electroweak scale, for as not to generate a new hierarchy problem in itself.


\enlargethispage{0.2cm}
We stress that goofy symmetry must be broken in some way, as the electroweak scale needs to be generated after all. Again, this bears some similarity to supersymmetry, which also has to be broken in some sequestered sector, and the breaking subsequently mediated to the Higgs to generate a viable EW scale. This supports the conclusion that also in the case of goofy protection of the EW scale, there is most likely a rich zoo of models for goofy symmetry breaking and mediation. This explains why we only briefly mention the spurion analysis, similar to soft supersymmetry breaking. 




For the case of imposing exotic space-time symmetries,
the spurion discussion enables an important observation about the ``intrinsic'' hierarchy problem.
The cut-off (or renormalization scale) itself in these cases has to transform like $\Lambda^2\mapsto-\Lambda^2$
in order to ensure the invariance of the effective action~\cite{Grzadkowski:2024, Ferreira:2025ate}, \footnote{We stress that the cutoff or renormalization scale does not transform because of goofy field transformations, but because of the additionally imposed goofy space-time 
transformation.}.
Curiously, this transforms exactly in such a way that it can compensate the goofy transformation of the Higgs bilinear, hence, give way to a goofy invariant contribution $\Lambda^2|H|^2$.

In absence of exotic space-time symmetries, one has to accept the fact that goofy transformations are necessarily explicitly (possibly softly) broken by the presence of (flavor universal, canonical) kinetic terms.
The goofy transformation is never imposed as an invariance of the full effective action, meaning also that the transformation behavior of a cut-off or renormalization scale is not dictated. Hence, scale generating spurions of various transformation behavior can be considered. 

The 2HDM examples in~\cite{Trautner:2025yxz} show that the mere presence of a mass scale does not mean that it automatically feeds back into all dimensionful (relevant) operators.  This explicitly shows that naive dimensional power counting alone is not good enough to
determine contribution of scales to other scales. What matters in addition is the representation under goofy transformations. 

In the SM EFT this demands for an additional power counting of scales, depending on their covariant goofy transformation behavior. In some sense this is analogous to helicity counting~\cite{Cheung:2015aba} for understanding the presence of ``holomorphic zeros'' in the (non-)renormalization of certain operators~\cite{Alonso:2014rga},\footnote{%
While this has not been shown, we strongly suspect that also helicity counting can be formulated based on an outer automorphism argument, just like for the case of goofy transformations~\cite{Trautner:2025yxz,deBoer:2025xx}.}. 
For example, scales generating effective operators of the type $|H|^{2+4n}$ require a violation of goofy symmetry and will give a boundary condition to feed into the Higgs mass at tree level, while scales that generate operators of the type $|H|^{4n}$ would -- if they actually enter the Higgs mass, which depends on details of the model -- give a boundary condition that is suppressed by the higher order, potentially sparse (since necessarily goofy breaking), radiative corrections~\footnote{As was pointed out to me by Stefan Antusch, also constraints from goofy transformation 
regarding expectation values of dimension-$4$ operators (dimension $0$ in the couplings), i.e.~vacuum energy, may be considered with potentially interesting conclusions about the cosmological constant problem, and this shall be explored in future work.}.

Lastly, we note that for non-trivially transforming fermion kinetic terms, the goofy transformation of the gauge coupling terms can be rolled over into a relative sign flip of certain gauge couplings. In two of our examples, this is the case exclusively for the electromagnetic and strong gauge couplings of the RH quarks. For flavor non-universal cases also inter-generational relative sign flips are possible. Currently we are not aware of any observable that is sensitive to either of such relative gauge coupling sign flips, but we think it would be very interesting to find some. Radiative corrections are organized according to such goofy sign flips (and outer automorphisms in general), irrespective of whether or not the according transformation is imposed as a symmetry.

\textit{Conclusions.--} Goofy transformations can be defined via their nature of non-trivially transforming \mbox{(gauge-)kinetic} terms. This implies that goofy transformations can prohibit bare scalar mass terms. This is the starting point for using goofy transformations to address the electroweak hierarchy problem.

Here we have shown several examples for goofy transformations that can prohibit the quadratic
term of the SM Higgs field, while permitting fermion Yukawa couplings. We found that there is a distinction necessary between the cases of goofy transformations which involve exotic space-time transformations, and cases which do not, which should also lead to important differences in their phenomenology.

On top of the discussion about the hierarchy problem, we have explicitly shown that goofy transformations can intrinsically be linked to the flavor structure and origin of CP violation. The continued exploration of goofy transformations, hence, should also include the non-trivial transformation of fermion Yukawa couplings, where goofy transformations could be a key missing ingredient in the search for solutions to the flavor puzzle.

The discrete goofy transformations discussed here appear closely related to some kind of discrete conformal transformation, in the sense that they reveal an important lesson when it comes to power counting of effective scales. While it is usually expected that every mass scale lifts a conformal protection, the existence of goofy transformations show that this is not necessarily the case. Heavy scales can exist but may not renormalize the Higgs quadratic term. In other cases, a Higgs mass counter term is prohibited but finite contributions from heavy scales can arise. This renders the Higgs mass calculable, as required for a true solution to the hierarchy problem. Phenomenological consequences of this kind of protection for the Higgs mass can only be explored on a model-by-model basis which certainly should be the goal of future work.
\section*{Acknowledgments}
This work is supported by the Portuguese Funda\c{c}\~ao para a Ci\^encia e a Tecnologia (FCT) through projects \href{https://doi.org/10.54499/2023.06787.CEECIND/CP2830/CT0005}{2023.06787.CEECIND}, \href{https://doi.org/10.54499/UID/00777/2025}{UID/00777/2025}, and contract \href{https://doi.org/10.54499/2024.01362.CERN}{2024.01362.CERN}, partially funded through POCTI (FEDER), COMPETE, QREN, PRR, and the EU.\\

\textbf{Note added.}-- While I was openly discussing the idea of goofy transformations for the hierarchy problem and consequences, Ref.~\cite{deBoer:2025jhc} appeared, which to some extent forced me to release my findings at the present stage. While ref.~\cite{deBoer:2025jhc} focuses on one case of a non-CP-type goofy transformation and imposes the presence of the additional space-time transformation, Eq.~\eqref{eq:goofy_st}, I also consider 
cases with no additional space-time transformations and CP-type transformations, and show that they necessarily have to have some level of flavor dependence. Where overlapping, our conclusions agree. Ref.~\cite{deBoer:2025jhc}, furthermore, provides some simple explicit extensions of the (goofy symmetric) SM that allows them to reintroduce the Higgs quadratic term and EW symmetry breaking. These are good examples for goofy-symmetric scenarios to address the EW hierarchy problem obeying the general arguments provided in my discussion section.
\bibliographystyle{utphys}
\bibliography{Bibliography}
\end{document}